\documentstyle[11pt]{article}

\newtheorem{Claim}{Claim}
\newtheorem{Lemma}{Lemma}
\newtheorem{Theorem}{Theorem}
\newtheorem{Definition}{Definition}

\newcommand{\comment}[1]{}

\def\proof{\noindent{\bf Proof:~}}
\newcommand{\qed}{\nobreak \ifvmode \relax \else
      \ifdim\lastskip<1.5em \hskip-\lastskip
      \hskip1.5em plus0em minus0.5em \fi \nobreak
      \vrule height0.75em width0.5em depth0.25em\fi}

\def\rket{\rangle}

\def\lbra{\langle}
\newcommand{\ket}[1]{| #1 \rangle}
\newcommand{\bra}[1]{\langle #1|}
\def\H{{\cal H}}
\def\eps{\epsilon}

\begin{document}
\title{On physical problems that are slightly more difficult than $QMA$}

\author{Andris Ambainis\\
University of Latvia and IAS, Princeton\\
Email: {\tt ambainis@lu.lv}}

\date{}

\maketitle

\begin{abstract}
We study the complexity of computational problems from quantum physics. 
Typically, they are studied using the complexity class $QMA$ (quantum counterpart of $NP$)
but some natural computational problems appear to be slightly harder than $QMA$.
We introduce new complexity classes consisting of problems that are solvable with a 
small number of queries to a $QMA$ oracle and use these complexity classes
to quantify the complexity of several natural computational problems (for example,
the complexity of estimating the spectral gap of a Hamiltonian).
\end{abstract}

\section{Introduction}

Quantum Hamiltonian complexity \cite{Osborne}
is a new field that combines quantum physics with computer 
science, by using the notions from computational complexity to study the complexity 
of problems that appear in quantum physics.

One of central notions of Hamiltonian complexity 
is the complexity class $QMA$ \cite{Knill,KSV,W,KKR,AN}
which is the quantum counterpart of NP. $QMA$ consists of 
all computational problems whose solutions can be verified in polynomial time
on a quantum computer, given a quantum witness (a quantum state on a polynomial
number of qubits).

$QMA$ captures the complexity of several interesting physical problems.
For example, estimating the ground state energy of a physical system (described by a
Hamiltonian) is a very important task in quantum physics. We can characterize
the complexity of this problem by showing that it is
$QMA$-complete, even if we restrict it to natural classes of Hamiltonians. 

One such natural restriction is to assume that the Hamiltonian is a sum of terms
in which each term is determined by interaction among at most $k$ particles, 
for some small $k$. Estimating the ground state energy of such a Hamiltonian is
known as the {\em k-local Hamiltonian problem} \cite{KR,KKR}. 
This problem is $QMA$-complete for any $k\geq 2$ \cite{KKR}.  
$QMA$-completeness also holds if we assume 
a natural geometric structure on the particles, with particles
arranged on a grid and each particle interacting with its nearest neighbours \cite{OT}
or restrict the Hamiltonian to certain natural interactions between qubits \cite{BL}.

$QMA$-completeness has been used to characterize the complexity of many computational problems
in quantum physics. (A number of other $QMA$-complete problems are given in \cite{Bookatz}.) 
But some natural physical problems seem to have a complexity that is slightly above $QMA$. 
For example, one such problem is estimating the spectral gap of a Hamiltonian $H$. 
The spectral gap of $H$ is the difference $\lambda_2-\lambda_1$ between the energy 
$\lambda_1$ of the ground state and the energy $\lambda_2$ of the state with the next smallest energy. To verify that $H$ has a spectral gap that is at least $\lambda$, 
one has to verify two statements:
\begin{enumerate}
\item[(a)]
The ground state energy $\lambda_1$ is at most $a$, for some $a$;
\item[(b)]
Any state that is orthogonal to the ground state has the energy at least $a+\lambda$.
\end{enumerate}
The first statement can be verified by a quantum algorithm V that takes the ground state 
$\ket{\psi}$ 
and estimates its energy. However, the second statement is hard to verify: a quantum algorithm can verify the existence of a state with energy at most $a+\lambda$ but not its non-existence!

In classical complexity theory, problems of similar nature are studied using generalizations of $NP$ such as:
\begin{itemize}
\item
$DP$, the class of ``differences" of two NP-complete problems, introduced 
by Papadimitriou and Yannakakis \cite{PY}. 
\item
The Boolean hierarchy, a sequence of complexity classes 
defined by taking intersections and unions of sets in $NP$ and $coNP$ \cite{CH,We,Wa,C+,C+1}.
\item
$P^{NP[log n]}$,
the class of problems that are solvable in polynomial time, if the algorithm is allowed to make
$O(\log n)$ queries to an $NP$ oracle \cite{BH,Wa90}.
\end{itemize}
There is a rich theory of such complexity classes 
and a number of natural computational problems have been shown to be complete for 
one of them (e.g., \cite{VV,PW,CM,HW,HHR,Krentel}).

In this paper, we connect this theory with quantum Hamiltonian complexity, by introducing $DQMA$ and $P^{QMA[log n]}$, the quantum counterparts of $DP$ and $P^{NP[log n]}$.
It turns out that these complexity classes are exactly the right tool for characterizing the complexity of natural computational problems (such as the spectral gap
problem mentioned above) in quantum physics! 

Namely, we have:
\begin{itemize}
\item
The problem of determining whether the ground state energy of a local Hamiltonian 
is close to a given value $a$ (as opposed to being substantially larger or substantially smaller) is $DQMA$-complete;
\item
The {\em ground state 
simulation problem}\cite{Osborne} in which we are given a Hamiltonian $H$ and an observable $A$
and have to distinguish whether the expectation of $A$ in the ground state of $H$ is at least $a+\epsilon$ or at most $a-\epsilon$ is $P^{QMA[log n]}$-complete.
\end{itemize}
Among these two problems, the second one is particularly interesting: 
determining the expectation of an observable
$A$ in a ground state of a Hamiltonian $H$ is important in many situations in quantum physics. 
It was known that this problem is $QMA$-hard \cite{Osborne}
but our result shows that it is probably harder than that
(unless $P^{QMA[log n]}=QMA$ which is unlikely).

For the problem of estimating the spectral gap of a Hamiltonian, 
we show that it is in $P^{QMA[log n]}$ and it is hard for a smaller complexity class,
$P^{UQMA[log n]}$ where queries to the $QMA$ oracle must be instances of $QMA$ with
either a unique witness or with no witness. It is not clear whether it is complete for any of these two classes.

Our results show that the complexity classes slightly above $QMA$ (which have not studied before) are quite useful for analyzing computational problems in quantum physics.
We expect that continuing this line of research could lead to other interesting
discoveries.

\comment{capture the complexity of several important
problems in quantum physics. Previously, it was known that those problems are $QMA$-hard but 
a more precise characterization was not known.
We expect that a further study of those complexity classes will lead to more interesting
results.}

\subsection{Related work} 
While $QMA$ has been studied in detail, there has been fairly little work on
generalizations of $QMA$ in directions similar to one that is considered in
this paper. 

The two main exceptions as as follows. 
First, Brown et al. \cite{BFS} and Shi and Zhang \cite{SZ} 
have studied the complexity class $\#QMA$ which is the quantum counterpart of $\#P$. 
The starting point for this work consists of two computational questions from quantum physics:
\begin{itemize}
\item
Determining how degenerate is the ground state of a Hamiltonian $H$ (determining the dimensionality of the eigenspace with the smallest eigenvalue);
\item
Density of states problem: determine the number of eigenvalues of $H$ in a given interval $[\lambda_1, \lambda_2]$ (allowing to miscount the eigenvalues 
in the intervals $[\lambda_1, \lambda_1+\frac{1}{n^c}]$ and $[\lambda_2-\frac{1}{n^c}, \lambda_2]$ around the endpoints of $[\lambda_1, \lambda_2]$).
\end{itemize}
Brown et al. \cite{BFS} show that both of those problems are complete for $\#QMA$
and that the class $\#QMA$ is equivalent to $\#P$ (i.e. $P^{\#P}=P^{\#QMA}$). 
The second result has been obtained independently by Shi and Zhang \cite{SZ}.

Our problem of estimating the spectral gap is related to the degeneracy 
problem in \cite{BFS}: if the ground state is degenerate, the spectral gap is 0.
The degeneracy problem is, however, much more general and, because of that, it has much higher complexity
($\#QMA = \#P$) than the problems in this paper (which are solvable with a small number of queries to a $QMA$ oracle).
As a result, the generalizations of $QMA$ in \cite{BFS} and in the current paper
are completely different. 

Second, Gharibian and Kempe \cite{GK} have studied a complexity class $cc-\Sigma_2$
which generalizes the classical class $\Sigma_2^p=NP^NP$. A problem is in 
$cc-\Sigma_2$ if, for any YES instance $x$, there exists a polynomial sized
classical proof $y$ such that for all polynomial size quantum proofs
$\ket{\psi}$, the verifier accepts $x, y$ and $\ket{\psi}$. Gharibian and Kempe
\cite{GK} then show $cc-\Sigma_2$ hardness for several quantum counterparts of
classical $\Sigma_2^p$-complete problems.

It is easy to see that $cc-\Sigma_2$ contains our class $DQMA$. The relation 
between $cc-\Sigma_2$ and $P^{QMA[log n]}$ is unclear: $cc-\Sigma_2$ can be
viewed as an $NP$ algorithm which is allowed to make one $QMA$ query. This is
stronger than $P^{QMA[log n]}$ in terms of the classical part ($NP$ instead of
$P$) but only allows one query to $QMA$.

Apart from \cite{BFS,SZ,GK}, we are 
not aware of any work that is more than distantly related. 
\cite{Koba,Aar,Harrow} and many others
have studied $QMA(k)$, a generalization of $QMA$ in which we are given several witness states $\ket{\psi_1}, \ldots, \ket{\psi_k}$
with a promise that they are not entangled. Both $QMA(k)$ and the complexity classes in 
the current paper are larger than $QMA$ but, apart from that, 
they do not seem to be related.

The complexity of spectral gap has also been studied in the context when the number of qubits grows to infinity 
(and the Hamiltonian is translationally invariant and, hence, can be described by a finite number of qubits) \cite{Cubitt}.
In this case, estimating the spectral gap becomes undecidable. This is somewhat similar to the undecidable tiling problems in which 
one has to decide whether it is possible to tile an infinite plane using a finite set of tiles \cite{Berger} and the proof of
undecidability of the spectral gap \cite{Cubitt} uses the undecidability of the tiling problem. 
The setting of this work is completely different from ours (in which every instance of the spectral gap has a fixed number of qubits $n$) and 
there is no relation between the results.

\section{Technical preliminaries}

\subsection{Notation}

We assume that we have a physical system consisting of $n$ qubits. 
The evolution of a physical system is
described by a {\em Hamiltonian} $H$ which is a Hermitian operator acting on the 
state-space of the system. If $\ket{\psi(t)}$ is the state of the system at time $t$,
then we have 
\[ \frac{d \ket{\psi(t)}}{dt} = H \ket{\psi(t)} .\]
In principle, any Hermitian $H$ can be a Hamiltonian of a physical system. On the other hand,
Hamiltonians of actual physical systems usually satisfy various locality constraints.

For example, a Hamiltonian may be formed by a combination of interactions, each of
which involves at most $k$ particles (qubits). Such Hamiltonians are called
$k$-{\em local}. More formally, $H$ is $k$-local if we can express it as 
$H=\sum_i H_i$, with each $H_i$ depending only on at most $k$ qubits. 
Throughout this paper, we assume that Hamiltonians $H$ are scaled so that 
all eigenvalues of $H$ are between 0 and $O(n^c)$ for some $c$ that is independent of $n$.

Mathematically, we can regard a Hamiltonian $H$ as a $D\times D$ Hermitian matrix.
The {\em ground state} of a Hamiltonian is just the eigenstate $\ket{\psi}$: $H\ket{\psi}=\lambda\ket{\psi}$
with the smallest eigenvalue $\lambda$. 
In physics terminology, the eigenvalue $\lambda$ is often called the {\em energy}
of the state $\ket{\psi}$. The degeneracy of the ground state is the dimension of the subspace
consisting of all $\ket{\psi}$: $H\ket{\psi}=\lambda\ket{\psi}$ with the smallest $\lambda$.

An {\em observable} is a Hermitian operator (which can be described by a Hermitian matrix) which
corresponds to a quantity of a physical system that can be measured. The value of
an observable $O$ on a state $\ket{\psi}$ is just $\bra{\psi} O \ket{\psi}$.
Often, observables are also $k$-local.

\subsection{Background on QMA}

We define the complexity class $QMA(c, s)$ to consist of all promise problems $L$ for which 
there exists a polynomial-time quantum algorithm $M(x, \ket{\psi})$ such that:
\begin{itemize}
\item
if $L(x)=1$, there exists $\ket{\psi}$ such that $M(x, \ket{\psi})$ outputs 1 with 
probability at least $c$;
\item
if $L(x)=0$, then for any $\ket{\psi}$, $M(x, \ket{\psi})$ outputs 1 with probability at most 
$s$.
\end{itemize}
We define $QMA=QMA(2/3, 1/3)$. It is known that the definition is robust 
w.r.t. choice of $c$ and $s$:

\begin{Theorem}
\cite{KSV}
$QMA=QMA(c, s)$ for any $c$ and $s$ such that $c\leq 1-2^{-p(x)}$, $s\geq 2^{-p(x)}$ and $c-s \geq \frac{1}{p(x)}$
for some polynomial $p(x)$.
\end{Theorem}

$QMA$ is sometimes called ``Quantum $NP$" and was first introduced by Kitaev \cite{Kitaev,KSV}
as a quantum counterpart of the classical complexity class $NP$. 
Instead of a classical witness and a classical verification algorithm
in the definition of $NP$, we have a quantum state $\ket{\psi}$ as a witness and a quantum 
algorithm for verifying this witness. Since the output of a quantum algorithm 
is probabilistic, it is natural to allow a small probability of error, making the 
definition of $QMA$ similar to the classical complexity class $MA$.

$QMA$ includes many natural computational problems from quantum physics.
A prototypical $QMA$-complete problem is 
$k$-LOCAL HAMILTONIAN$(H, a, b)$. 
In this problem, we are given a Hamiltonian $H$ (which can be expressed as $H=\sum_i H_i$
with each $i$ being $k$-local for a constant $k$) acting on $n$ qubits 
and real numbers $a, b$: $b\geq a + \frac{1}{n^c}$ (where $c$ is a fixed constant). 
The task is to distinguish between the two cases:
\begin{itemize}
\item
$k$-LOCAL HAMILTONIAN$(H, a, b)= 0$: the ground state energy of $H$ is at least $b$;
\item
$k$-LOCAL HAMILTONIAN$(H, a, b)= 1$: the ground state energy of $H$ is at most $a$
\end{itemize}
under a promise that one of those two cases occurs.

We can always modify the Hamiltonian $H$ so that it has a state with energy exactly $b$.
Therefore, we can modify the promise to ``the ground state energy is either at most $a$ or
is exactly $b$".

\begin{Theorem}
\cite{KKR}
\label{thm:kkr}
2-LOCAL HAMILTONIAN is $QMA$-complete. 
\end{Theorem}

{\bf Closure properties}  
$QMA$ satisfies closure properties that are similar to the closure properties of NP.
Let $L_1$ and $L_2$ be two promise problems. We define $L=L_1\wedge L_2$ by
$L(x)=L_1(x)~AND~L_2(x)$ and $L=L_1\vee L_2$ by
$L(x)=L_1(x)~OR~L_2(x)$. (In both cases, if one of $L_1(x)$ and $L_2(x)$ is undefined,
$L(x)$ is undefined as well.) It is easy to show

\begin{Theorem}
\label{thm:closure}
If $L_1\in QMA$, $L_2\in QMA$, then $L_1\wedge L_2\in QMA$ and $L_1\vee L_2\in QMA$.
\end{Theorem}

\subsection{QMA with unique witnesses}

$UQMA$ is a variant of $QMA$ in which we require the quantum witness $\ket{\psi}$ to be unique in the $L(x)=1$ case.
It is non-trivial to define when a quantum witness is ``unique" because the space of all witnesses $\ket{\psi}$ is continuous.
Therefore, if $M(x, \ket{\psi})$ outputs 1 with a probability $p>2/3$, then $M(x, \ket{\psi'})$ will also output 1 with probability
at least 2/3 whenever $\ket{\psi'}$ is sufficiently close to $\ket{\psi}$.

The solution to this problem is as follows. We say that a quantum witness $\ket{\psi}$ is unique if $M$ rejects any $\ket{\psi'}\perp\ket{\psi}$ with
a high probability. Then, the only witnesses $\ket{\phi}$ that are accepted are the ones that have a sufficiently high overlap with $\ket{\psi}$. 

\begin{Definition}\cite{Aharonov+}
The complexity class $UQMA(c, s)$ consists of all promise problems $L$ for which 
there exists a polynomial-time quantum algorithm $M(x, \ket{\psi})$ such that:
\begin{itemize}
\item
if $L(x)=1$, there exists $\ket{\psi}$ such that 
\begin{enumerate}
\item[(a)]
$M(x, \ket{\psi})$ outputs 1 with probability at least $c$;
\item[(b)]
If $\ket{\psi'}\perp \ket{\psi}$, then $M(x, \ket{\psi'})$ 
outputs 1 with probability at most $s$.
\end{enumerate}
\item
if $L(x)=0$, then for any $\ket{\psi}$, $M(x, \ket{\psi})$ outputs 1 with probability at most 
$s$.
\end{itemize}
\end{Definition}

Classically, we can reduce general instances of problems in $NP$ to instances with a unique witness (this is the well-known Valiant-Vazirani theorem \cite{VV}). 
It is not known whether a similar result is true in the quantum case \cite{Aharonov+,J+}.

An example of a problem in $UQMA$ is UNIQUE $k$-LOCAL HAMILTONIAN$(H, a, b)$ in
which $H, a, b$ are similar to $k$-LOCAL HAMILTONIAN and
the task is to distinguish between the two cases:
\begin{itemize}
\item
UNIQUE $k$-LOCAL HAMILTONIAN$(H, a, b)= 0$: the ground state energy 
of $H$ is at least $b$;
\item
UNIQUE $k$-LOCAL HAMILTONIAN$(H, a, b)= 1$: the ground state energy 
of $H$ is at most $a$ and any other eigenstate of $H$ has the energy at least $b$.
\end{itemize}
under a promise that one of those two cases occurs.

$UQMA$-complete problems have not been studied before but we can show that this problem is complete for $UQMA$, 
similarly to how $k$-LOCAL HAMILTONIAN is $QMA$-complete.

\begin{Theorem}
UNIQUE 3-LOCAL HAMILTONIAN is $UQMA$-complete. 
\end{Theorem}

The theorem follows by adapting the proof that 3-LOCAL HAMILTONIAN is $QMA$-complete by Kempe and Regev \cite{KR}.
We include it in appendix \ref{app:uqma}.

It is plausible that the proof of $QMA$-completeness of 2-LOCAL HAMILTONIAN by Kitaev, Kempe and Regev \cite{KKR} can be adapted
to show that UNIQUE 2-LOCAL HAMILTONIAN is $UQMA$-complete but we have not verified that.

\section{Our results}

\subsection{Problems slightly beyond QMA}

We now consider three problems whose true complexity seems to be slightly beyond QMA. 
\begin{enumerate}
\item
Given a Hamiltonian $H$, is it true that its ground state energy is close to a given 
number $a$?

\begin{Definition}
EXACT $k$-LOCAL HAMILTONIAN$(H, a, \eps, \delta)$. 

Given a Hamiltonian $H$ and real numbers $a, \eps, \delta$: $\eps\geq \frac{1}{n^c}$, 
$\delta\geq \frac{1}{n^c}$, 
distinguish between the following two cases:
\begin{itemize}
\item
EXACT $k$-LOCAL HAMILTONIAN$(H, a, \eps, \delta)= 0$: the ground state 
energy of $H$ is in the interval $[a-\eps, a+\eps]$;
\item
EXACT $k$-LOCAL HAMILTONIAN$(H, a, \eps, \delta)=1$: 
the ground state energy of $H$ does not belong to the interval $[a-\eps-\delta, a+\eps+\delta]$.
\end{itemize}
\end{Definition}
\item
Given a Hamiltonian $H$, estimate its spectral gap: 
the difference $\lambda_2-\lambda_1$ where $\lambda_1$ and $\lambda_2$ are the 
two smallest eigenvalues of $H$.  

The spectral gap $\lambda_2-\lambda_1$ is an important physical quantity
in several contexts. In the context of quantum computing, it is related to the running time of adiabatic quantum algorithms \cite{Farhi} which roughly scales as 
$\frac{1}{(\lambda_2-\lambda_1)^2}$. To estimate the running time, we are interested in 
distinguishing whether the spectral gap is small (close to 0) or large. We can formalize this as

\begin{Definition}
SPECTRAL GAP$(H, \eps)$.

Given a Hamiltonian $H$ and a real number $\eps\geq \frac{1}{n^c}$,
distinguish between two cases: 
\begin{itemize}
\item
SPECTRAL GAP$(H, \eps)= 1$: $\lambda_2-\lambda_1\leq \eps$;
\item
SPECTRAL GAP$(H, \eps)= 0$: $\lambda_2-\lambda_1\geq 2 \eps$.
\end{itemize}
\end{Definition}
\item
In the ground state simulation problem \cite{Osborne}, we are given a Hamiltonian $H$
(we assume that $H$ is $k$-local) and a physical quantity described by 
an observable $A$ (we also assume that $A$ is 
$k$-local). The task is to estimate $\lbra \psi |A| \psi\rket$ where $\ket{\psi}$ is the
ground state of $H$. We can turn this into a (promise) decision problem in a standard way:
we define that the task is to output 1 if $\lbra \psi |A| \psi\rket \leq \alpha_1$
and to output 0 if $\lbra \psi |A| \psi\rket \geq \alpha_2$ for some $\alpha_1, \alpha_2$:
$\alpha_1<\alpha_2$, $\alpha_2-\alpha_1=\Omega(1/n^c)$. 

We can study this problem in two versions: exact or approximate:

\begin{Definition} 
EXACT-SIMULATION$(H, A, \alpha_1, \alpha_2)$.

Given a Hamiltonian $H$, an observable $A$ and numbers $\alpha_1, \alpha_2$
with $\alpha_2-\alpha_1\geq \frac{1}{n^c}$ (where $n$ is the input size),
distinguish between the following two cases:
\begin{itemize}
\item
EXACT-SIMULATION$=1$ if $H$ has a ground state $\ket{\psi}$ with $\lbra \psi |A| \psi\rket \leq \alpha_1$;
\item
EXACT-SIMULATION$=0$ if $H$ has no ground state 
$\ket{\psi}$ with $\lbra \psi |A| \psi\rket \leq \alpha_2$.
\end{itemize}
\end{Definition}

\begin{Definition}
APPROX-SIMULATION$(H, A, \alpha_1, \alpha_2, \eps)$.

Given a Hamiltonian $H$, an observable $A$ and numbers $\alpha_1, \alpha_2, \eps$
with $\alpha_2-\alpha_1\geq \frac{1}{n^c}$, $\eps\geq \frac{1}{n^c}$,
distinguish between the following two cases:
\begin{itemize}
\item
APPROX-SIMULATION$=1$ if $H$ 
has a ground state $\ket{\psi}$ with $\lbra \psi |A| \psi\rket \leq \alpha_1$;
\item
APPROX-SIMULATION$=0$ if, for any $\ket{\psi}$ with $\bra{\psi} H \ket{\psi}\leq 
\lambda+\eps$ (where $\lambda$ is the smallest eigenvalue of $H$), 
we have $\bra{\psi} A \ket{\psi}\geq \alpha_2$.
\end{itemize}
\end{Definition}

In this paper, we study APPROX-SIMULATION because it is more similar in spirit to the 
other problems that we consider.
We also think that it may be more natural because it is more robust w.r.t. small perturbations in
the Hamiltonian $H$.
\end{enumerate}

For all of these 3 problems, to verify that $P=1$ we must verify a combination 
of a statement that involves existence of a quantum state with certain properties
with a statement that involves non-existence of a quantum state. 
For example, for EXACT $k$-LOCAL HAMILTONIAN, we have to verify that 
\begin{enumerate}
\item[(a)]
There exists a state $\ket{\psi}$ such that $\lbra \psi | H |\psi\rket \leq a+\eps$;
\item[(b)]
There is no state $\ket{\psi}$ such that $\lbra \psi | H |\psi\rket \leq a-\eps-\delta$.
\end{enumerate}
The first statement can be verified in $QMA$ but the second statement is an
opposite of what can be verified in $QMA$.

\subsection{Complexity results}
\label{sec:results}

In classical complexity theory \cite[Chapter 17.1]{P}, such problems are characterized 
using complexity classes that are slightly above $NP$. 
One such class is $DP$ \cite{PY} which consists of all languages
$L$ such that $L=L_1\cap L_2$, $L_1\in NP$, $L_2\in coNP$. Examples of
problems belonging to $DP$ are:
\begin{itemize}
\item
EXACT TSP: we are given an instance of traveling salesman problem (TSP) and
have to determine if the shortest TSP tour has the length exactly $k$;
\item
UNIQUE SAT: we have to determine whether a SAT formula has exactly one 
satisfying assignment
\item
CRITICAL SAT: we have to determine whether it is true that a SAT formula is unsatisfiable
but removing an arbitrary clause from it would result in a satisfiable formula.
\end{itemize}
In all of those cases, it is easy to show that the problem is in $DP$, by expressing 
it as a combination of two statements, one of which can be verified in $NP$ and the 
other is a negation of a statement that can be verified in $NP$.
Interestingly, all 3 of those problems are also $DP$-complete
(as shown by Papadimitriou and Yannakakis \cite{PY}, Valiant and Vazirani \cite{VV}
and Papadimitriou and Wolfe \cite{PW}, respectively).

We can characterize the complexity of EXACT $k$-LOCAL HAMILTONIAN by
a new quantum complexity class $DQMA$ (which is a quantum counterpart of $DP$).

\begin{Definition}
$DQMA$ is a class consisting of all promise problems $L$ for which 
we have $L_1, L_2\in QMA$ such that:
\begin{itemize}
\item
If $L(x)=1$ then $L_1(x)=1$ and $L_2(x)=0$;
\item
If $L(x)=0$ then $L_1(x)$ and $L_2(x)$ are both defined and either $L_1(x)=0$ or $L_2(x)=1$. 
\end{itemize}
\end{Definition}

\begin{Theorem}
\label{thm:1}
EXACT 3-local HAMILTONIAN is $DQMA$-complete.
\end{Theorem}

\proof
In section \ref{sec:proof1}.
\qed

The other two problems (APPROX-SIMULATION and SPECTRAL GAP) are more difficult. 
If we are given a $QMA$ oracle, we 
can solve APPROX-SIMULATION with $O(\log n)$ queries to the oracle,
in a following way:
\begin{enumerate}
\item
We use the $QMA$ oracle and binary search to obtain an estimate $a$ for $\lambda_1$ 
(the smallest eigenvalue of $H$) such that $\lambda_1\in[a, a+\eps/2]$;
\item
We use one more query to the $QMA$ oracle to verify the statement: ``there exists 
$\ket{\psi}$ which is a linear combination of eigenvectors of $H$ with eigenvalues
in $[a, a+\eps/2]$ and satisfies $\bra{\psi} A \ket{\psi} \leq \alpha_1$".
\end{enumerate}
For the first step, we need $O(\log \frac{1}{\eps} ) = O(\log n)$ queries 
to obtain an estimate $a$ with a sufficient precision. The second step requires 1 query. 

This shows that APPROX-SIMULATION belongs to a complexity 
class $P^{QMA[log n]}$ in which a polynomial
time classical algorithm $M$ is allowed to make $O(\log n)$ queries to an
oracle solving a promise problem in $QMA$. 
APPROX-SIMULATION is also complete for this complexity class.

\begin{Theorem}
\label{thm:apx}
APPROX-SIMULATION is $P^{QMA[log n]}$-complete.
\end{Theorem}

\proof
In section \ref{sec:apx}.
\qed

SPECTRAL GAP also belongs to $P^{QMA[log n]}$ (by a similar binary search argument) 
but it is not clear whether it is
$P^{QMA[log n]}$-complete. 

The reason why it is difficult to show $P^{QMA[log n]}$-hardness
of SPECTRAL GAP is as follows. We assume that we are trying to embed a computation
consisting of $O(\log n)$ queries to a $QMA$ oracle into one instance of 
SPECTRAL GAP. We can assume that the queries are to an oracle solving
$k$-LOCAL HAMILTONIAN problem. Then, it could be the case that the Hamiltonians $H$ 
in the queries have very small spectral gaps (of the order smaller than $1/n^c$ for any fixed $c$). 
In this case, it is difficult to expect that the Hamiltonian for SPECTRAL GAP 
obtained by combining them would have a larger spectral gap of order $\Omega(1/n^c)$,
as required in the case when SPECTRAL GAP=0.

If this problem does not arise (i.e., if all queries are to instances of UNIQUE $k$-LOCAL HAMILTONIAN), 
we can embed a computation involving $O(\log n)$ queries to a $QMA$ oracle into an instance of SPECTRAL GAP.
Since UNIQUE $k$-LOCAL HAMILTONIAN is $UQMA$-complete, this gives us

\begin{Theorem}
\label{thm:spec}
\begin{enumerate}
\item[(a)]
SPECTRAL GAP $\in P^{QMA[log n]}$;
\item[(b)] 
SPECTRAL GAP, for $O(\log n)$-local Hamiltonians, is $P^{UQMA[log n]}$-hard.
\end{enumerate}
\end{Theorem}

\proof
In section \ref{sec:spec}
\qed

We note that SPECTRAL GAP is probably not in $P^{UQMA[log n]}$, for the following reason.
Let $H$ be the Hamiltonian that is the input for the SPECTRAL GAP problem.
If the spectral gap of $H$ is small, then it is likely that the query Hamiltonians (which are
produced from $H$) will also have a small spectral gap and, thus, they will not be instances
of a $UQMA$ problem.

\section{Conclusion}

In this paper, we have connected complexity classes defined using a small number of queries to an $NP$ oracle
with quantum Hamiltonian complexity, by introducing $DQMA$ and $P^{QMA[log n]}$, the quantum counterparts of $DP$ and $P^{NP[log n]}$.
We then used the new complexity classes to characterize the complexity of several natural computational problems (such as simulation problem and spectral gap) 
in quantum physics.

Some of the problems that we study have been known to be
$QMA$-hard but not in $QMA$. Yet, the possibility of capturing the complexity of these
problems via complexity classes slightly above $QMA$ was not noticed before.

We think that this is just the beginning for a new research area
further work in this direction can lead to other interesting discoveries.
Some specific open questions resulting from our work are:
\begin{enumerate}
\item
Can we quantify the complexity of SPECTRAL GAP more precisely?
\item
What can we prove about the complexity of EXACT-SIMULATION?
Intuitively, it should be much harder than APPROX-SIMULATION 
because very small changes to the Hamiltonian $H$
can change an instance with EXACT-SIMULATION$(H)=1$ into an instance with 
EXACT-SIMULATION$(H)=0$.
\item
Our hardness results use 3-local Hamiltonians for EXACT $k$-LOCAL HAMILTONIAN
and $O(\log n)$-local Hamiltonians for APPROX-SIMULATION and SPECTRAL-GAP.

Since most of Hamiltonians which actually occur in nature obey quite strong locality constraints
(typically, they are $k$-local for quite small constant $k$), it would be interesting to
know whether one can achieve similar hardness results using $k$-local Hamiltonians for smaller $k$.
\end{enumerate}
More general topics for future research are:
\begin{enumerate}
\item
Quantifying the complexity of other physical problems through the complexity classes
$DQMA$, $P^{QMA[log n]}$ and other similar complexity classes;
\item
Developing a quantum theory of classes ``slightly above $QMA$",
along the lines of the classical theory of classes ``slightly above $NP$".
\end{enumerate}

{\bf Acknowledgment.}
This research was inspired by discussions with Sergei Bravyi and Norbert Schuch during
the programme ``Entanglement and correlations in many-body quantum mechanics" at Erwin Schrodinger Institute in Vienna, Austria.
The authors thanks the referees for CCC'2014 whose comments helped to improve the paper.

The author was supported by FP7 FET projects QCS and QALGO and ERC Advanced Grant MQC (at the University of Latvia) and by National Science 
Foundation under agreement No. DMS-1128155 (at IAS, Princeton). 
Any opinions, findings and conclusions or recommendations expressed in this material are those of the author(s) and 
do not necessarily reflect the views of the National Science Foundation.

\appendix

\section{Proofs of our main results}

\subsection{Complexity of EXACT HAMILTONIAN}
\label{sec:proof1}

In this section, we prove Theorem \ref{thm:1}.

To show that EXACT $k$-local HAMILTONIAN $\in DQMA$, we observe that 
computing $E=~$EXACT $k$-local HAMILTONIAN$(H, a, \eps, \delta)$ reduces to computing 
$E_1=$ $k$-local HAMILTONIAN$(H, a+\eps, a+\eps+\delta)$
and $E_2=$ $k$-local HAMILTONIAN$(H, a-\eps-\delta, a-\eps)$. 
If $E=1$, then $E_1=1$ and $E_2=0$. If $E=0$, then both $E_1$ and $E_2$ are defined
and either $E_1=0$ or $E_2=1$.

To show the completeness, let $L\in DQMA$ and $L_1$, $L_2$ be the corresponding problems
from $QMA$. By Theorem \ref{thm:kkr}, we can reduce both $L_1$ and $L_2$ to 
2-LOCAL HAMILTONIAN with the same $a$ and $b$ in both cases.
We also assume that $a=\eps$, $b=2\eps$.

Let $H_1$ and $H_2$ be the two instances of 2-local Hamiltonian problem
produced by our reduction from $L_1(x)$ and $L_2(x)$.
We assume that both $H_1$ and $H_2$ are Hamiltonians on $m$ qubits and
define an $m+1$ qubit Hamiltonian 
\[ H = \ket{0}\bra{0} \otimes H_1 + 3 \ket{1}\bra{1} \otimes H_2 + 
4\eps \ket{0}\bra{0} \otimes I  .\]

We claim that $L(x)$ is equivalent to EXACT 3-LOCAL HAMILTONIAN$(H, 4.5\eps, \eps/2, \eps)$.
Let $\H_0$ ($\H_1$) be the subspace consisting of all states with the first qubit being 
$\ket{0}$ ($\ket{1}$) and let $\lambda_0$, $\lambda_1$ be the lowest energies of $H_1$ and 
$H_2$. Then, the lowest energy state of $H$ on the subspace $\H_0$ has the energy 
$\lambda_0+4\eps$ and the lowest energy state on $\H_1$ has the energy $3\lambda_1$.

We consider three cases:
\begin{enumerate}
\item
$L(x)=1$. $\lambda_0\in[0, \eps]$ and
$\lambda_0+4\eps \in[4\eps, 5\eps]$.
The lowest energy state on the subspace $\H_0$ has the energy at least 
$3\lambda_1\geq 6\eps$ (since $L_2(x)=0$).
\item
$L(x)=0$ and $L_2(x)=1$. 
Then, the lowest energy state on $\H_1$ has the energy 
$3\lambda_1\leq 3\eps$ (since $L_2(x)=1$).
\item
$L(x)=0$ and $L_1(x)=L_2(x)=0$. 
Then, the lowest energy state on $\H_0$ has the energy 
$\lambda_1+4\eps \geq 6\eps$ (since $L_2(x)=0$).
and the lowest energy state on $\H_1$ has the energy at least 
$6\eps$ (similarly to the first case). \qed
\end{enumerate}

\subsection{Complexity of APPROX-SIMULATION}
\label{sec:apx}

In this section, we prove Theorem \ref{thm:apx}. 

{\bf Part 1:} APPROX-SIMULATION$\in P^{QMA[log n]}$. 
As we already described in section \ref{sec:results}, the algorithm consists of two steps:
\begin{enumerate}
\item
We use the $QMA$ oracle and binary search to obtain an estimate $a$ for $\lambda$ 
(the smallest eigenvalue of $H$) such that $\lambda\in[a, a+\eps/2]$;
\item
We use one more query to the $QMA$ oracle to verify the statement: ``there exists 
$\ket{\psi}$ which is a linear combination of eigenvectors of $H$ with eigenvalues
in $[a, a+\eps/2]$ and satisfies $\bra{\psi} A \ket{\psi} \leq \alpha_1$".
\end{enumerate}
The first step is performed as follows:
\begin{enumerate}
\item
Let $\delta=\eps/2$.
\item
Start with $[a, b]=[0, 1]$.
\item
As long as $b-a> \frac{\eps}{2}$, repeat:
\begin{enumerate}
\item
Query $k$-LOCAL HAMILTONIAN($H, \frac{a+b}{2}-\frac{\delta}{4}, \frac{a+b}{2}+\frac{\delta}{4}$).
\item
Depending on the answer, set $[a, b]=[a, \frac{a+b}{2}+\frac{\eps}{4}]$ or
$[a, b]=[\frac{a+b}{2}-\frac{\eps}{4}, b]$.
\end{enumerate}
\end{enumerate}  
Each repetition decreases the size of the interval $[a, b]$ by almost a half and,
after $O(\log \frac{1}{\eps})=O(\log n)$ repetitions, we have $b-a\leq \frac{\eps}{2}$.

For the second step, we query the $QMA$ oracle whether there exists a state $\ket{\psi}$ which is accepted (with a high probability)
by a following quantum algorithm $M$. (Formally, this can be done by reducing the existence of such $\ket{\psi}$ to 
an instance of $k$-LOCAL HAMILTONIAN and querying the oracle for $k$-LOCAL HAMILTONIAN.)

let $\H$ be the Hilbert space on which $H$ acts. The input space of $M$ is $(\H)^{\otimes k}$ for sufficiently 
large $k=poly(n)$. $M$ first performs eigenvalue estimation for operator $H$ on each copy of $\H$ with precision $\eps/4$ (and sufficiently small error). If at least one of estimates for eigenvalues is more than $a+\frac{3\eps}{4}$, $M$ outputs 0. Otherwise, it uses $k$ copies
of $\H$ to estimate the average of $\bra{\psi} A\ket{\psi}$ over all $k$ registers
with a precision $\delta < \frac{\alpha_2-\alpha_1}{2}$. 
(Since $\delta=\Omega(\frac{1}{n^c})$, such precision
can be achieved using $k=poly(n)$ copies.) 
If the resulting estimate is at most $\alpha_1+\delta$,
$M$ outputs 1. Otherwise, $M$ outputs 1.

If APPROX-SIMULATION=1, then the smallest eigenvalue of $H$ is
$\lambda\leq a+\frac{\eps}{2}$ and the corresponding eigenvector $\ket{\psi}$
satisfies $\bra{\psi} A \ket{\psi} \leq \alpha_1$.
Then, inputting $\ket{\psi}^{\otimes k}$ to $M$ results in $M$ outputting 1 with a high probability. 

If APPROX-SIMULATION=0, we would like to show that there is no $\ket{\psi}\in (\H)^{\otimes k}$
for which $M$ outputs 1 with a substantial probability. 
We first note that any $\ket{\psi}\in (\H)^{\otimes k}$ can be expressed as a linear combination
of $\ket{\psi_1}\otimes \ket{\psi_2} \otimes \ldots \otimes \ket{\psi_k}$
where each $\ket{\psi_i}$ is an eigenstate of $H$.
We express $\ket{\psi}=\ket{\psi_+}+\ket{\psi_-}$, with 
$\ket{\psi_+}$ being the part of $\ket{\psi}$ consisting of 
$\ket{\psi_1}\otimes \ket{\psi_2} \otimes \ldots \otimes \ket{\psi_k}$
where each $\ket{\psi_i}$ has the the eigenvalue that is at most $a+\eps$
and $\ket{\psi_-}$ consisting of all other 
$\ket{\psi_1}\otimes \ket{\psi_2} \otimes \ldots \otimes \ket{\psi_k}$.

On $\ket{\psi_-}$, the eigenvalue estimation part of $H$ results in $M$ outputting 0 with a high probability.
Conditional on $\ket{\psi}$ not being rejected, the remaining state is close to $\ket{\psi_{+}}$.
However, $\bra{\psi_{+}}A\ket{\psi_{+}}\geq \alpha_2$ for any $\ket{\psi_{+}}$ that is a linear combination of 
eigenvectors of $H$ with eigenvalues at most $\lambda+\eps>a+\eps$.
Therefore, estimating the average of $\bra{\psi_{+}}A\ket{\psi_{+}}$ results in an estimate
that is at least $\alpha_2-\delta$, with a high probability (and the same happens when, instead of 
$\ket{\psi_+}$, we estimate $\bra{\psi}A\ket{\psi}$ for a state $\ket{\psi}\approx \ket{\psi_+}$.
In this case, $M$ also outputs 0.


{\bf Part 2:} APPROX-SIMULATION is $P^{QMA[log n]}$-hard. 
By Theorem \ref{thm:kkr}, we assume that queries are to an
oracle $O$ for 2-LOCAL HAMILTONIAN problem, with $a=\eps$ 
and $b=3\eps$ where $\eps=1/n^c$. 

Let $M$ be a polynomial time classical algorithm that makes $O(\log n)$ queries to
a $QMA$ oracle.
Given an input $x$, we can simulate $M(x)$ for all possible combinations of answers 
by oracle $O$. Let $d$ be the maximum number of queries made by $M(x)$. 
Then, we have $d \leq c \log n$ for some $c$. Therefore, there are $O(n^c)$
possible sequences of answers and this simulation runs in a polynomial time.
Let $H^{(i)}_{y_1\ldots y_{i-1}}$ be the Hamiltonian that is asked by $M(x)$ in the 
$i^{\rm th}$ query, if the answers to the previous queries are $y_1, \ldots, y_{i-1}$.
We can assume that all of those Hamiltonians act on the same Hilbert space $\H$
consisting of the same number of qubits.
Let $ans_{y_1\ldots y_d}$ be the answer output by $M(x)$, 
if the answers to $M$'s queries are $y_1\ldots y_d$. 

We take a bigger Hilbert space $\H'=\otimes_{i=1}^d \H_{i, 1} \otimes \H_{i, 2}$
where $\H_{i, 1}$ is the Hilbert space for one qubit and $\H_{i, 2}$ is
isomorphic to $\H$. We let the Hamiltonians $H^{(i)}_{y_1\ldots y_{i-1}}$
act on $\H_{i, 2}$. We consider the Hamiltonian 
\[ H_t = \sum_{i=1}^t \frac{1}{4^{i-1}} \sum_{y_1\ldots y_{i-1}}  \otimes_{j=1}^{i-1} \ket{y_j}\bra{y_j}_{\H_{j, 1}} 
\otimes \left( 2\eps \ket{0}\bra{0}_{\H_{i, 1}} + 
\ket{1}\bra{1}_{\H_{i, 1}} \otimes H^{(i)}_{y_1\ldots y_{i-1}} \right) .\]

\begin{Claim}
\label{cl:inductive}
Let $y_1\ldots y_d$ be the correct answers to queries made by $M(x)$.
The ground state of $H_t$ is in the subspace 
\[ \H_{y_1\ldots y_t} = \otimes_{i=1}^t \ket{y_i}\bra{y_i} \otimes \H_{i, 2} .\]
Let $\lambda_t$ be the ground state energy of $H_t$. Then, energy of 
any state in $\H_{y'_1\ldots y'_t}$, for any $y'_1\ldots y'_t\neq y_1\ldots y_t$,
is at least $\lambda_t+\frac{\eps}{4^{t-1}}$.
\end{Claim}

This claim reduces the problem solved by $M$ to 
APPROX-SIMULATION in a following way. 
We take the Hamiltonian $H_d$ and define the observable $A$ as the sum of 
$\otimes_{i=1}^d \ket{y_i}\bra{y_i}_{\H_{i, 1}}$ over all $y_1\ldots y_d$ 
such that $M(x)$ outputs 1 if the answers to queries are $y_1, \ldots, y_d$.
Then, $M(x)=1$ is equivalent to 
APPROX-SIMULATION$(H_d, A, 0, 1, \eps/4^{d-1})=1$.

\proof [of Claim]
By induction over $t$. We assume that the claim is true for $H_{t-1}$ and prove 
that it is also true for $H_t$. For each term of the form
\[ \otimes_{j=1}^{i-1} \ket{y_j}\bra{y_j}_{\H_{j, 1}} 
\otimes \left( 2\eps \ket{0}\bra{0}_{\H_{i, 1}} + 
\ket{1}\bra{1}_{\H_{i, 1}} H^{(i)}_{y_1\ldots y_{i-1}} \right), \]
its ground state has the energy between 0 and $3\eps$.
Therefore, the lowest energy of a state in $\H_{y_1\ldots y_{t-1}}$ is
at most 
\[ E = \lambda_{t-1}+\frac{1}{4^{t-1}} \cdot 3 \eps \]
and the lowest energy of a state in any other $\H_{y'_1\ldots y'_{t-1}}$
is at least 
\[ \lambda_{t-1}+ \frac{\eps}{4^{t-2}}  =
E + \frac{\eps}{4^{t-1}} .\]
This means that the ground state of $H_t$ must lie in
$\H_{y_1\ldots y_{t-1}}$. On this subspace, $H_t$ acts in the same way 
as 
\begin{equation}
\label{eq:two} 
H_{t-1} + \frac{1}{4^{t-1}} \left( 2\eps \ket{0}\bra{0}_{\H_{i, 1}} + 
\ket{1}\bra{1}_{\H_{i, 1}} H^{(i)}_{y_1\ldots y_{i-1}} \right) 
\end{equation}
Since the two terms of (\ref{eq:two}) act on different qubits, the ground state
is the tensor product of their ground states. We have two cases:
\begin{enumerate}
\item
If $y_t=0$, the ground state of the second term is any state that has $\ket{0}$ in
$\H_{i, 1}$, with the energy $2\eps$. Therefore, the overall ground state
is in $\H_{y_1\ldots y_{t-1}0}$, with the energy 
$\lambda_t=\lambda_{t-1}+\frac{2\eps}{4^{t-1}}$.
Any state in $\H_{y_1\ldots y_{t-1}1}$ must have energy at least 
\[ \lambda_{t-1}+\frac{3\eps}{4^{t-1}}  = \lambda_t + \frac{\eps}{4^{t-1}} .\]
\item
If $y_t=1$, the ground state of the second term is a state of the form $\ket{1}\otimes\ket{\psi}$, with $\ket{\psi}$ being the ground state of 
$ H^{(i)}_{y_1\ldots y_{i-1}}$. This state has energy at most $a=\eps$.
Hence, the ground state of $H_t$ is in $\H_{y_1\ldots y_{t-1}0}$, with energy 
$\lambda_t\leq \lambda_{t-1}+\frac{1}{4^{t-1}}\eps$.
Any state in $\H_{y_1\ldots y_{t-1}1}$ must have energy at least 
\[ \lambda_{t-1}+\frac{2\eps}{4^{t-1}}  = \lambda_t + \frac{\eps}{4^{t-1}}.\]
\end{enumerate}
\qed
\qed

\subsection{Complexity of SPECTRAL GAP}
\label{sec:spec}

In this section, we prove Theorem \ref{thm:spec}.

(a)
SPECTRAL GAP can be solved by a following algorithm:
\begin{enumerate}
\item
Use binary search with $O(\log n)$ queries to 
the QMA oracle to determine an interval $[a, a+\eps/4]$
such that $\lambda \in [a, a+\eps/4]$ where $\lambda$
is the smallest eigenvalue of $H$.
\item
Use the QMA oracle to determine if there exists a quantum state $\ket{\psi}$
accepted by the following algorithm $M$:
\begin{enumerate}
\item
The input to $M$ is a quantum state in $\H'=\H\otimes \H$ where $\H$ is
the Hilbert space on which $H$ acts.
\item
Let $\H_{-}$ by the antisymmetric subspace of $\H'$ (the subspace spanned by 
the states of the form $\ket{\varphi}\otimes\ket{\varphi'} - 
\ket{\varphi'}\otimes\ket{\varphi}$). 
\item
$M$ measures whether the input state $\ket{\psi}$
belongs to $\H_{-}$ or the subspace $\H^{\perp}$ which is perpendicular to
$\H_{-}$ and rejects if the answer is $\H^{\perp}$.
\item
If the answer is $\H_{-}$, $M$ performs eigenvalue estimation for 
$H\otimes H$ on the state $\ket{\psi}$, with precision $\eps/5$.
$M$ outputs 1 if the estimate for the eigenvalue is at most 
$2a+\frac{7}{4}\epsilon$ and 0 otherwise.
\end{enumerate}
\end{enumerate}
To analyze the algorithm, we first observe that, restricted to $\H_-$,
eigenstates of $H\otimes H$ are of the form 
\begin{equation}
\label{eq:form} 
\ket{\psi_i}\otimes\ket{\psi_j} - 
\ket{\psi_j}\otimes\ket{\psi_i}
\end{equation}
where $\ket{\psi_i}$ and $\ket{\psi_j}$ are
eigenstates of $H$.

If the spectral gap of $H$ is at most $\epsilon$, let $\ket{\psi_1}$
and $\ket{\psi_2}$ be two eigenstates with the smallest eigenvalues 
$\lambda_1$ and $\lambda_2$. Then, $\ket{\psi_1}\otimes\ket{\psi_2} - 
\ket{\psi_2}\otimes\ket{\psi_1}$ is an eigenstate of $H\otimes H$ 
with an eigenvalue
\[ \lambda_1+\lambda_2 \leq \left(a+\frac{\eps}{4}\right)+
\left(a+\frac{\eps}{4}\right)+ \epsilon = 2a + \frac{3}{2} \epsilon .\]
If the spectral gap of $H$ is $2\eps$ or more, let
$\ket{\psi}$ be an eigenstate of $H\otimes H$. 
Then, $\ket{\psi}$ is of the form (\ref{eq:form}).
Let $\lambda_i$ and $\lambda_j$ be the eigenvalues 
of $\ket{\psi_i}$ and $\ket{\psi_j}$.
Then, the eigenvalue of $\ket{\psi}$ is 
\[ \lambda_i+\lambda_j \geq a + (a+2\eps) = 2a + 2\eps .\]
In both cases, estimating the eigenvalue with precision $\eps/5$ 
will give the right answer.

(b) We assume that queries are to an
oracle $O$ for 2-LOCAL HAMILTONIAN problem, with $a=\eps$ 
and $b=3\eps$ where $\eps=1/n^c$, with a promise that the spectral
gap of the Hamiltonians that are being queried is at least $\eps$. 

Without a loss of generality, we assume that $M(x)$ always makes
the maximum number of questions $d$ to the oracle $O$.
Similarly to the proof of Theorem \ref{thm:apx}, we 
simulate $M(x)$ for all possible combinations of answers by $O$ and let
$H^{(i)}_{y_1\ldots y_{i-1}}$ be the Hamiltonian that is asked by $M(x)$ in the 
$i^{\rm th}$ query, if the answers to the previous queries are $y_1, \ldots, y_{i-1}$.

Let $H_0$ be any fixed Hamiltonian (on $\H$) with the following properties:
\begin{itemize}
\item
$H_0$ has a unique ground state with an eigenvalue $2\eps$;
\item
All other eigenvalues of $H_0$ are at least $3\eps$. 
\end{itemize} 
We build a sequence of Hamiltonians $H_1, \ldots, H_d$ where 
\[ H_t = \sum_{i=1}^t \frac{1}{4^{i-1}} \sum_{y_1\ldots y_{i-1}}  
\otimes_{j=1}^{i-1} \ket{y_j}\bra{y_j}_{\H_{j, 1}} 
\otimes \left( \ket{0}\bra{0}_{\H_{i, 1}} \otimes (H_0)_{\H_{i, 2}} \right.\]
\[ \left. + 
\ket{1}\bra{1}_{\H_{i, 1}} \otimes H^{(i)}_{y_1\ldots y_{i-1}} \right) .\]
Similarly to Claim \ref{cl:inductive}, 
the ground state of $H_t$ lies in the subspace $\H_{y_1\ldots y_t}$.
Moreover, we have
\begin{Claim}
As a Hamiltonian on $\otimes_{i=1}^t \H_{i, 1}\otimes\H_{i, 2}$, $H_t$
has a unique ground state and spectral gap of at least $\frac{\eps}{4^t}$.
\end{Claim}

\proof
By induction. We assume that $H_{t-1}$ satisfies the claim and show that
this assumption implies that $H_t$ also satisfies the claim. (The base case for
$H_1$ follows by slightly modifying the proof of the inductive case.)

The eigenstates of $H_t$ can be expressed as 
$\ket{\psi}\otimes\ket{y}\otimes\ket{\phi}$ where
\begin{itemize}
\item
$\ket{\psi}\in\otimes_{i=1}^t \H_{i, 1}\otimes\H_{i, 2}$ is an eigenstate of $H_{t-1}$;
\item
$y\in\{0, 1\}$;
\item
$\ket{\phi}$ is an eigenstate of $H_0$ (if $y=0$) or 
$H^{(i)}_{y_1\ldots y_{i-1}}$ (if $y=1$).
\end{itemize}
The eigenvalue of this state is $\lambda_{t-1}+\Delta$ where $\lambda_{t-1}$ is the eigenvalue
of $\ket{\psi}$ (as an eigenstate of $H_{t-1}$) and $\Delta$ is the eigenvalue of 
$\ket{\phi}$ (as an eigenstate of $
\frac{1}{4^t}H_0$ or $\frac{1}{4^t}H^{(i)}_{y_1\ldots y_{i-1}}$).
To minimize this, $\lambda_{t-1}$ and $\Delta$ must both be the smallest eigenvalues of
the respective Hamiltonians. Let $\ket{\psi}\otimes \ket{y} \otimes\ket{\phi}$
be the corresponding eigenvector.

Let $\ket{\psi'}\otimes \ket{y'} \otimes\ket{\phi'}$
be any other eigenstate of $H_{t-1}$.
If $\ket{\psi'}\neq\ket{\psi}$, then $\lambda_t$ is larger by
at least $\frac{\eps}{4^{t-1}}$ (from the inductive assumption).
If $\ket{\psi}=\ket{\psi'}$, 
we have two cases:
\begin{enumerate}
\item
If $y=y'$, the eigenvalue $\Delta$ for $\ket{\phi'}$
is larger than $\Delta$ for $\ket{\phi}$
by at least $\frac{\eps}{4^{t}}$ (for $y=0$, this is true because we choose a Hamiltonian
with the spectral gap $\eps$ as $H_0$; for $y=1$, it follows from the promise about the spectral gap of the Hamiltonians that we are querying). 
\item
If $y\neq y'$, then one of $\ket{\phi}$, $\ket{\phi'}$ is an eigenvector of $H_0$ and the other is an eigenvector of $H^{(i)}_{y_1\ldots y_{i-1}}$. Since the smallest eigenvalue of $H_0$ is $2\eps$ and the smallest eigenvalue of 
$H^{(i)}_{y_1\ldots y_{i-1}}$ is either at most $a=\eps$ or at least
$b=3\eps$, this results in a difference of at least $\frac{\eps}{4^t}$ between the 
corresponding eigenvalues $\Delta$.
\end{enumerate}
\qed

We now define 
\[ H' = \sum_{y_1\ldots y_d} \otimes_{i=1}^d \ket{y_i}\bra{y_i}_{\H_{i, 1}} \]
with the summation over all $y_1, \ldots, y_d$ such that $M(x)$ outputs 0 
if the answers to queries are equal to $y_1, \ldots, y_d$.
We then add an extra qubit $B$ to the system and define
\[ H_{final} = I_B \otimes H_d + \eps \ket{0}\bra{0}_B\otimes H' .\]
We claim that $M(x)=1$ is equivalent to SPECTRAL GAP$(H_{final}, \eps/4^d)$:
\begin{enumerate}
\item
If $M(x)=1$, the spectral gap is 0 because $H_{final}$ has 
2 orthogonal states with the smallest eigenvalue: $\ket{0}\otimes\ket{\psi}$
and $\ket{1}\otimes\ket{\psi}$ where $\ket{\psi}$ is the ground state of $H$; 
\item
If $M(x)=0$, the state with the smallest eigenvalue is $\ket{1}\otimes\ket{\psi}$.
Its eigenvalue differs from the eigenvalue of $\ket{0}\otimes\ket{\psi}$ by $\eps$
(because of the $\ket{0}\bra{0}_B\otimes H'$ term in $H_{final}$) and from 
any other eigenvalue by at least $\frac{\eps}{4^d}$ (because of the spectral gap of
$H_d$). 
\end{enumerate}
\qed

\section{$UQMA$-completeness of UNIQUE $k$-LOCAL HAMILTONIAN}
\label{app:uqma}

\subsection{Background on $QMA$-completeness of LOCAL HAMILTONIAN}

We can reduce $L$ to 3-LOCAL HAMILTONIAN using 
the reduction of Kitaev and Regev \cite{KR}. We claim that this is actually 
a reduction to UNIQUE 3-LOCAL HAMILTONIAN. 

This reduction works as follows. We first amplify the success probability of the
verifier $M$ to $1-\frac{1}{2^n}$, using the error reduction for $QMA$/$UQMA$ described in
the next subsection. We then
represent the verifier circuit $M$ as a sequence of quantum
gates acting on 1 or 2 qubits: $U_1$, $U_2$, $\ldots$, $U_T$. We can then construct a reduction
of $L$ to $O(\log n)$-LOCAL HAMILTONIAN. To do that, we introduce an $\lceil \log_2 (T+1) \rceil$ 
qubit register $C$ with basis states $\ket{0}, \ldots, \ket{T}$.
Let $A_i$ (for $i=1, \ldots, m$) be the ancilla qubits (which must be initialized to $\ket{0}$ 
at the beginning) and $O$ be the output qubit of $M$. We define a Hamiltonian
\[ H_1 = H_{in} + H_{out} + H_{prop} \]
where 
\[ H_{in} = \sum_{i=1}^m \ket{1}_{A_i} \otimes \ket{0}_C, \]
\[ H_{out} = \ket{0}_O \otimes \ket{T}_C, \]
\[ H_{prop} = \sum_{t=0}^{T-1} H_{prop, i}, \]
\[ H_{prop, i} = \frac{1}{2} \left( I \otimes \ket{t}\bra{t} + I \otimes \ket{t-1}\bra{t-1} -
U_t \otimes \ket{t}\bra{t-1} - U^{\dagger}_t \otimes \ket{t-1}\bra{t} \right) .\]
To obtain a reduction to 3-LOCAL HAMILTONIAN, we use the unary representation for $C$, 
representing $\ket{i}_C$ as
\begin{equation}
\label{eq:unary} 
\ket{\underbrace{0\ldots 0}_i \underbrace{1\ldots 1}_{T-i}} 
\end{equation}
and choose $H$ as
\[ H_2 = H'_{in} + H'_{out} + H'_{prop} + H_{clock} \]
where 
\[ H_{clock} = T^{6} \sum_{i=1}^{T-1} \ket{1}\bra{1}_i \otimes \ket{0}\bra{0}_{i+1} \] 
is a Hamiltonian that penalizes the states of $C$ that are not valid unary representations and
$H'_{in}$, $H'_{out}$, $H'_{prop}$ are 3-local Hamiltonians that satisfy the following 
requirements:
\begin{itemize}
\item 
$H_{in} = \Pi H'_{in} \Pi$, $H_{out} = \Pi H'_{out} \Pi$, $H_{prop} = \Pi H'_{prop} \Pi$
where $\Pi$ is the projection to the subspace $\H_{legal}$ consisting of the states 
in which $C$ is in one of valid unary states of form (\ref{eq:unary});
\item
$\|H'_{in}+H'_{out}+H'_{prop}\|=O(T)$. 
\end{itemize}
Then, we have

\begin{Theorem}
\cite{KR}\begin{enumerate}
\item
If $M$ accepts $\ket{\psi}\otimes \ket{0^m}$ for some $\ket{\psi}$ 
with probability more than $1-\epsilon$, $H_2$
has an eigenvalue that is smaller than $\frac{\epsilon}{T+1}$.
\item
If $M$ accepts $\ket{\psi}\otimes \ket{0^m}$ with any $\ket{\psi}$ 
with probability at most than $\epsilon$ on any $\ket{\psi}$, $H_2$
has no eigenvalue that is smaller than $\frac{c}{T^3}$ for some constant $c>0$.
\end{enumerate}
\end{Theorem}

To show that this is also a reduction from $L\in UQMA$ to UNIQUE 3-LOCAL HAMILTONIAN,
we need to show 

\begin{Lemma}
\label{lem:main}
Assume that there is a state $\ket{\psi}$ such that $M$ accepts $\ket{\psi}\otimes 
\ket{0}^{\otimes m}$ with probability at least $1-\epsilon$ and accepts any $\ket{\phi}\otimes 
\ket{0}^{\otimes m}$, $\ket{\phi}\perp\ket{\psi}$ with probability at most $\epsilon$.
Then, the second smallest eigenvalue of $H_2$ is at least $\frac{c}{T^3}$
for an appropriately chosen $c>0$.
\end{Lemma}

\subsection{Error reduction}

Let $L\in UQMA$. Because of the error reduction for $UQMA$ (Theorem 2 of \cite{J+} which builds on
a similar result for QMA by Mariott and Watrous) we can build a verifier $M$ with the following
properties:
\begin{itemize}
\item
If $L(x)=1$, there is a state $\ket{\psi}$ such that $M$ accepts $\ket{\psi}\otimes 
\ket{0}^{\otimes m}$ with probability at least $1-\epsilon$ and accepts any $\ket{\phi}\otimes 
\ket{0}^{\otimes m}$, $\ket{\phi}\perp\ket{\psi}$ with probability at most $\epsilon$;
\item
If $L(x)=0$, $M$ accepts any $\ket{\phi}\otimes \ket{0}^{\otimes m}$ 
with probability at most $\epsilon$;
\end{itemize}
where $\epsilon = \frac{1}{2^n}$ and both the number of ancilla qubits $m$ and the running time
of the verifier $T$ are $poly(n)$. 
We observe that this implies the following. 

\begin{Claim}
\label{cl:2}
Let $\ket{\psi_1}\perp\ket{\psi_2}$. Then, for at 
least one of $i\in\{1, 2\}$, $M$ accepts $\ket{\psi_i}\otimes \ket{0}^{\otimes m}$
with probability at most $\frac{(1+\sqrt{\epsilon})^2}{2}$.
\end{Claim}

\proof
Since $\ket{\psi_1}\perp\ket{\psi_2}$, the angle between 
one of $\ket{\psi_i}$ and $\ket{\psi}$ is at least $\frac{\pi}{4}$. Therefore,
\[ \ket{\psi_i} = \alpha \ket{\psi} + \beta \ket{\psi^{\perp}} \]
where $\ket{\psi^{\perp}}\perp\ket{\psi}$, $|\alpha|\leq \frac{1}{\sqrt{2}}$.
Since $\ket{\psi}$ is accepted with probability at most 1 and $\ket{\psi^{\perp}}$
with probability at most $\epsilon$, the claim follows.
\qed


\subsection{Useful facts from linear algebra}

We use two facts from linear algebra:

\begin{Lemma}
\label{lem:1}
\cite{HJ}[Corollary 7.7.4]
Let $H, H'$ be Hermitian matrices such that $H\succeq H'$\footnote{In this paper, 
$A\succeq B$ means that $A-B$ is positive semidefinite.}.
Let $\lambda_1 \geq \lambda_2 \geq \ldots$ 
and $\lambda'_1\geq \lambda'_2\geq \ldots$ be the eigenvalues
of $H$ and $H'$, respectively.
Then, $\lambda_i \geq \lambda'_i$ for all $i$.
\end{Lemma}

\begin{Lemma}
\label{lem:2}
Let $H=\Pi+\Pi'$ where 
$\Pi$ and $\Pi'$ are two projections acting on the same Hilbert space $\H$.
Then, we can decompose $\H$ into a direct sum of 1-dimensional and 
2-dimensional subspaces $\H_i$ such that each $H(\H_i)\subseteq \H_i$, with
\begin{enumerate}
\item[(a)]
each 1-dimensional subspace spanned by $\ket{\psi}$ that is an eigenvector of
both $\Pi$ and $\Pi'$;
\item[(b)]
each 2-dimensional subspace spanned by two eigenvectors of
$\Pi$, one with eigenvalue 1 and one with eigenvalue 0 (and, similarly, by
two eigenvectors of $\Pi'$).
\end{enumerate}
\end{Lemma}

A consequence of this lemma is that we can determine the smallest eigenvalue of $H$
by looking at each subspace separately. Each one dimensional subspace $\H_i$ is spanned
by one eigenstate $\ket{\psi_i}$. Therefore, we either have $\Pi\ket{\psi_i} = \ket{\psi_i}$
or $\Pi\ket{\psi_i} = 0$ (and similarly for $\Pi'$). Hence, the eigenvalue of
$\ket{\psi_i}$ is 0, 1 or 2.

For two-dimensional subspaces $\H_i$, each of them is spanned by two eigenstates
$\ket{\psi_{i, 1}}$ and $\ket{\psi_{i, 2}}$ of $\Pi$ and two eigenstates
$\ket{\phi_{i, 1}}$ and $\ket{\phi_{i, 2}}$ of $\Pi'$.
If both $\ket{\psi_{i, 1}}$ and $\ket{\psi_{i, 2}}$ have the same eigenvalue,
then $\H_i$ decomposes into a sum of two one-dimensional subspaces, spanned
by $\ket{\phi_{i, 1}}$ and $\ket{\phi_{i, 2}}$.
Hence, we can assume that $\Pi \ket{\psi_{i, 1}}=\ket{\psi_{i, 1}}$ and 
$\Pi\ket{\psi_{i, 2}}=0$. Similarly, $\Pi' \ket{\phi_{i, 1}}=\ket{\phi_{i, 1}}$ and 
$\Pi'\ket{\phi_{i, 2}}=0$. We assume that the phases of $\ket{\phi_{i, j}}$ 
have been chosen so that $\lbra \psi_{i, 1}|\phi_{i, 1}\rket$ is a non-negative real. 
Then, the eigenstates of $H=\Pi+\Pi'$ on $\H_i$ are  
\[ \ket{\varphi_{i, 1}} = \ket{\psi_{i, 1}} + \ket{\phi_{i, 1}} \]
and $\ket{\varphi_{i, 2}}\perp\ket{\varphi_{i, 1}}$,
with eigenvalues $2\cos^2 \frac{\alpha}{2}$ and $2\sin^2 \frac{\alpha}{2}$
where $\alpha$ is the angle between $\ket{\psi_{i, 1}}$ and $\ket{\phi_{i, 1}}$.
Since $\lbra \psi_{i, 1}|\phi_{i, 1}\rket$ is a non-negative real,
we have $\alpha\in[0, \frac{\pi}{2}]$ and, hence,
$\cos^2 \frac{\alpha}{2}\geq \frac{1}{2} \geq \sin^2 \frac{\alpha}{2}$

\subsection{Analysis of Hamiltonian $H_1$}

We first analyze the Hamiltonian $H_1$. 
The counterpart of Lemma \ref{lem:main} is

\begin{Lemma}
Assume that there is a state $\ket{\psi}$ such that $M$ accepts $\ket{\psi}\otimes 
\ket{0}^{\otimes m}$ with probability at least $1-\epsilon$ and accepts any $\ket{\phi}\otimes 
\ket{0}^{\otimes m}$, $\ket{\phi}\perp\ket{\psi}$ with probability at most $\epsilon$.
Then, the second smallest eigenvalue of $H_1$ is at least $\frac{c}{T^3}$
for an appropriately chosen $c>0$.
\end{Lemma}

\proof
We express $H_1 = H_{prop} + H_{other}$ where $H_{other} = H_{in} + H_{out}$.
The smallest non-zero eigenvalue of $H_{other}$ is 1 because all terms of $H_{other}$ are diagonal in the computational basis and the smallest non-zero eigenvalue of each term is 1.
The smallest non-zero eigenvalue of $H_{prop}$ is $\lambda=\Omega(1/T^2)$ \cite{KR}.
By Lemma \ref{lem:1}, we can replace $H_1$ with $\lambda (\Pi_{other} + \Pi_{prop})$
where $\Pi_{other}$ and $\Pi_{prop}$ are projections to the subspaces spanned by
all eigenvectors of $H_{other}$ and $H_{prop}$ with non-zero eigenvalues.

Let $\delta$ be the second smallest eigenvalue of $H' = \Pi_{other}+\Pi_{prop}$.
We will show that $\delta>\frac{c}{T+1}$ for some constant $c$. This implies that the 
second smallest eigenvalue of $H_1$ is at least $c\lambda = \Omega(1/T^3)$.

The zero eigenspace of $\Pi_{prop}$ consists of all the ``history states" of the form
\begin{equation}
\label{eq:history} \frac{1}{\sqrt{T+1}} \sum_{i=0}^T \ket{\psi_i} \otimes \ket{i} 
\end{equation}
where $\ket{\psi_i}$ is the state of the verifier $M$ after $i$ steps
(given that the starting state is $\ket{\psi_0}$).

We consider the decomposition of $H' = \Pi_{other}+\Pi_{prop}$ given by Lemma \ref{lem:2}.
We claim that:
\begin{enumerate}
\item[(a)]
Each subspace $\H_i$ contains at most one eigenvector of $H'$ with an eigenvalue less than 1.
\item[(b)]
If $\H_i$ contains an eigenvector with eigenvalue at most
$\delta<1$, it also contains a history state
such that 
\begin{equation}
\label{eq:cond} 
\|\Pi_{in} \ket{\psi_0} + \Pi_{out} \ket{\psi_T}\|\leq \sqrt{2(T+1)\delta} .
\end{equation} 
\end{enumerate}
 
If $\H_i$ is one-dimensional, the first part is obvious. For the second part, let $\ket{\psi}\in \H_i$. If $\H_i$ is one-dimensional, then $\ket{\psi}$ must be an eigenvector
of both $\Pi_{other}$ and $\Pi_{prop}$ and, if the eigenvalue of $\ket{\psi}$ is at most
$\delta<1$, then $\Pi_{other}\ket{\psi}=\Pi_{prop}\ket{\psi} = 0$.
$\Pi_{prop} \ket{\psi}=0$ means that $\ket{\psi}$ is a ``history state" of the form
(\ref{eq:history}). $\Pi_{other}\ket{\psi}=0$ means that 
$\Pi_{in}\ket{\psi}=\Pi_{out}\ket{\psi}=0$, implying that (\ref{eq:cond}) is satisfied.

If $\H_i$ is 2-dimensional, it must contain one eigenvector of $\Pi_{prop}$ with
eigenvalue 1 and one eigenvector of $\Pi_{other}$ with eigenvalue 1.
If $\lambda_1$ and $\lambda_2$ are the eigenvalues of $H'$ for the two eigenvectors 
in $\H_i$, then
\[ \lambda_1+\lambda_2 = Tr (H' \Pi_{\H_i}) = Tr(\Pi_{prop} \Pi_{\H_i})+
Tr(\Pi_{other} \Pi_{\H_i}) = 2.\]
This means that at most one of $\lambda_1$, $\lambda_2$ can be less than 1. 
For the part (b), let $\ket{\psi_1}$ and $\ket{\psi_2}$ be the 
1-eigenvectors of $\Pi_{prop}$ and $\Pi_{other}$ in $\H_i$.
We assume that the phases of $\ket{\psi_1}$ and $\ket{\psi_2}$ have been chosen so that
$\lbra \psi_1 \ket{\psi_2}$ is a positive real. Then, the eigenvectors of $H'$ are
$\ket{\psi_1}+\ket{\psi_2}$ and $\ket{\psi_1}-\ket{\psi_2}$ 
with eigenvalues $2\cos^2 \frac{\beta}{2}$ and $2\sin^2 \frac{\beta}{2}$ where
$\beta$ is such that $\cos \beta = \lbra \psi_1 \ket{\psi_2}$.
The smallest of those eigenvalues is $2\sin^2 \frac{\beta}{2}$.

We now express the left hand side of (\ref{eq:cond}) through $\beta$.
The history state $\ket{\psi_{hist}}$ contained in $\H_i$ is a 0-eigenstate
of $\Pi_{prop}$. Therefore, it is orthogonal to $\ket{\psi_1}$. Since all those 
states lie in a 2-dimensional subspace $\H_i$, this means that $\lbra \psi_2 \ket{\psi_{hist}}=\sin\beta$.
$\lbra \psi_2 \ket{\psi_{hist}}$ is also the length of the projection of
$\ket{\psi_{hist}}$ to $\ket{\psi_2}$. This projection is the same as 
\[ \Pi_{other} \ket{\psi_{hist}} = \frac{1}{\sqrt{T+1}}\Pi_{in} \ket{\psi_0} 
+ \frac{1}{\sqrt{T+1}}\Pi_{out} \ket{\psi_T}.\]
Therefore, we have $\|\Pi_{in} \ket{\psi_0} + \Pi_{out} \ket{\psi_T}\|=\sin\beta$. If
$2\sin^2 \frac{\beta}{2} \leq \delta$, then
\[ \sin\beta = 2\sin\frac{\beta}{2} \cos\frac{\beta}{2} \leq 
2 \sqrt{\frac{\delta}{2}(1-\frac{\delta}{2})} \leq \sqrt{2\delta} .\]
Hence, if the 2nd smallest eigenvalue of $H'$ is $\delta$, then we have two
orthogonal history states $\ket{\psi_{hist, 1}}$ and $\ket{\psi_{hist, 2}}$
satisfying the condition (\ref{eq:cond}).

For each of those history states, we define a new history state
$\ket{\psi'_{hist, i}}$ in the following way.
Let $\ket{\psi_{i, 0}}$ denote the starting state $\ket{\psi_0}$ that is used to define
$\ket{\psi_{hist, i}}$. Since $\ket{\psi_{hist, 1}}$ and $\ket{\psi_{hist, 2}}$ are
orthogonal, $\ket{\psi_{1, 0}}$ and $\ket{\psi_{2, 0}}$ must be orthogonal, as well.  
Let
\[ \ket{\psi'_{i, 0}} = \frac{\ket{\psi_{i, 0}}- \Pi_{in} \ket{\psi_{i, 0}}}{
\| \ket{\psi_{i, 0}}- \Pi_{in} \ket{\psi_{i, 0}} \|} .\]
Because of (\ref{eq:cond}), we have 
\begin{equation}
\label{eq:7} 
\| \ket{\psi_{i, 0}}-\ket{\psi'_{i, 0}} \| \leq 
\sqrt{2(T+1)\delta} + o(\sqrt{(T+1)\delta}) .
\end{equation}
Let $\alpha$ be the angle between $\ket{\psi'_{1, 0}}$ and $\ket{\psi'_{2, 0}}$.
Because of $\ket{\psi_{1, 0}}\perp\ket{\psi_{2, 0}}$ and (\ref{eq:7}), we have
$\alpha\leq 2\sqrt{2(T+1)\delta}+o(\sqrt{(T+1)\delta})$. 
We take the plane spanned by $\ket{\psi'_{i, 0}}$ for $i\in\{1, 2\}$ and in this plane 
choose $\ket{\psi''_{i, 0}}$ so that the angle between $\ket{\psi'_{i, 0}}$ and 
$\ket{\psi''_{i, 0}}$ is $\frac{\pi}{4}-\frac{\alpha}{2}$ and
$\ket{\psi''_{1, 0}}\perp \ket{\psi''_{2, 0}}$. Then,
\begin{equation}
\label{eq:8} 
 \| \ket{\psi''_{i, 0}}-\ket{\psi'_{i, 0}} \| 
\leq \sqrt{2(T+1)\delta} + o(\sqrt{(T+1)\delta}) .
\end{equation}
We have $\Pi_{in} \ket{\psi''_{i, 0}}=0$ (because $\Pi_{in} \ket{\psi'_{i, 0}}=0$ for
both $i\in\{1, 2\}$ and $\ket{\psi''_{i, 0}}$ are in the plane spanned by 
$\ket{\psi'_{i, 0}}$).  
Let $\ket{\psi_{i, T}}$ ($\ket{\psi''_{i, T}}$)
be the final states of the computation if we start it in
the state $\ket{\psi_{i, 0}}$ ($\ket{\psi''_{i, 0}}$). 
Because of (\ref{eq:7}), (\ref{eq:8}) and 
unitary transformations being length-preserving, we have 
\begin{equation}
\label{eq:9} 
 \| \ket{\psi_{i, T}}-\ket{\psi''_{i, T}} \| 
\leq 2\sqrt{2(T+1)\delta} +  o(\sqrt{(T+1)\delta}) .
\end{equation} 
Also, because of (\ref{eq:cond}), we have 
$\|\Pi_{out} \ket{\psi_{i, T}}\|\leq \sqrt{2(T+1)\delta}$.
Together with (\ref{eq:9}), this means 
\[ \|\Pi_{out} \ket{\psi''_{i, T}}\|
\leq 3\sqrt{2(T+1)\delta} + o(\sqrt{(T+1)\delta}) .\]
Moreover, if we have some other starting state $\ket{\psi_0}$ in the
plane spanned by $\ket{\psi''_{i, 0}}$, then
\[ \ket{\psi_0} = \alpha \ket{\psi''_{1, 0}} +
\beta  \ket{\psi''_{2, 0}} .\]
For the corresponding final state $\ket{\psi_T}$, we have
\[ \|\Pi_{out} \ket{\psi_T}\|
\leq (|\alpha|+|\beta|) 3\sqrt{2(T+1)\delta} + o(\sqrt{(T+1)\delta}) \]
\[ \leq 6 \sqrt{(T+1)\delta} + o(\sqrt{(T+1)\delta}) .\]
Hence, any state $\ket{\psi_0}$ in the 
plane spanned by $\ket{\psi''_{i, 0}}$ is accepted with
probability at least $1-36(T+1)\delta-o((T+1)\delta)$. 
Because of Claim \ref{cl:2}, we must have
$(T+1)\delta\geq c$ for some constant $c>0$.
Then, $\delta\geq \frac{c}{T+1}$.
\qed

\subsection{Analysis of Hamiltonian $H_2$}

We now complete the proof of Lemma \ref{lem:main}. Similarly to the previous case, we show that, if the Hamiltonian $H_2$ has two eigenstates with small eigenvalues, then there must be a two-dimensional subspace such that the verifier accepts any state in this subspace with a 
probability close to 1. 

We express $H_2=H'_{clock} + H'_{other}$ where $H'_{other} = H'_{in} + H'_{out} + H'_{prop}$. 
Let $\ket{\psi_1}, \ket{\psi_2}$ be two eigenstates of $H_2$ with eigenvalues 
$\lambda_i$ less than $\frac{\delta}{T^3}$. We will prove that
$\delta>c$ for an appropriately chosen $c>0$.

For each $i\in\{1, 2\}$, we express
\[ \ket{\psi_i} = \ket{\psi_{i, 1}} + \ket{\psi_{i, 2}} \]
with $\ket{\psi_{i, 1}} \in \H_{legal}$, $\ket{\psi_{i, 2}}\perp \H_{legal}$.
Because of
\[ \lambda_i = \bra{\psi_i} H_2 \ket{\psi_i} \geq
\bra{\psi_i} H_{clock} \ket{\psi_i} = T^6 \|\psi_i\|^2 \]
must have $\|\psi_{i, 2}\|^2\leq \frac{\delta}{T^{9}}$.
 
Since $\bra{\psi_i} H'_{other} \ket{\psi_i} \leq \frac{\delta}{T^3}$,
$\|\psi_{i, 2}\|^2\leq \frac{\delta}{T^{9}}$ and $\|H'_{other}\| = O(T)$,
we have 
\[ \bra{\psi_{i, 1}} H'_{other} \ket{\psi_{i, 1}} \leq \frac{\delta+o(1)}{T^3} .\]
Moreover, since $\|\psi_{i, 2}\|^2\leq \frac{\delta}{T^{9}}$, we have 
$\| \psi_{i, 1} \| \geq 1- \sqrt{\frac{\delta}{T^{9}}}$ and 
$|\lbra \psi_{1, 1} \ket{\psi_{2, 1}}| = O(\frac{1}{T^{9/2}})$.
Hence, we can replace $\ket{\psi_{i, 1}}$ by $\ket{\psi'_{i, 1}}$ such that 
$\|\psi'_{i, 1}\| = 1$ and $\ket{\psi'_{1, 1}}\perp \ket{\psi'_{2, 1}}$ and we
still have
\[ \bra{\psi'_{i, 1}} H'_{other} \ket{\psi'_{i, 1}} \leq \frac{\delta+o(1)}{T^3} .\]
Since $\ket{\psi'_{i, 1}}\in \H_{legal}$, we have
\[ \bra{\psi'_{i, 1}} H'_{other} \ket{\psi'_{i, 1}} =
\bra{\psi'_{i, 1}} \Pi_{legal} H'_{other} \Pi_{legal} \ket{\psi'_{i, 1}} 
 .\]
This means that the Hamiltonian $\Pi_{legal} H'_{other} \Pi_{legal}$ has two eigenvalues
that are at most $\frac{\delta+o(1)}{T^3}$.

We now recall that, on the subspace $\H_{legal}$, terms of $\Pi_{legal} H'_{other} \Pi_{legal}$
act in the same way as the terms of $H_1$. Therefore, we can now use the proof from the previous subsection.
\qed

\end{document}